\documentclass[pra,twocolumn,superscriptaddress,showpacs,floatfix]{revtex4-1}
\usepackage{graphicx,amssymb}
\usepackage[fleqn]{amsmath}
\usepackage{amsfonts}
\usepackage{dcolumn}
\usepackage{bm}
\usepackage{braket}
\usepackage{mathtools}
\usepackage{multirow} 

\newcommand{\gras}[1]{\boldsymbol{#1}}

\begin{document}

\preprint{ }

\title{Bound states of dipolar molecules studied with the Berggren expansion method}

 \author{K. Fossez}
\affiliation{Grand Acc\'el\'erateur National d'Ions Lourds (GANIL), CEA/DSM - CNRS/IN2P3,
BP 55027, F-14076 Caen Cedex, France}%

\author{N. Michel}
\affiliation{Department of Physics \&
Astronomy, University of Tennessee, Knoxville, Tennessee 37996, USA}
\affiliation{Physics Division, Oak Ridge National Laboratory, Oak Ridge, Tennessee 37831, USA}
 
\author{W. Nazarewicz}
\affiliation{Department of Physics \&
Astronomy, University of Tennessee, Knoxville, Tennessee 37996, USA}
\affiliation{Physics Division, Oak Ridge National Laboratory, Oak Ridge, Tennessee 37831, USA}
\affiliation{Institute of Theoretical Physics, University of Warsaw, ul. Ho\.za
69, 00-681 Warsaw, Poland}

\author{M. P{\l}oszajczak}
\affiliation{
Grand Acc\'el\'erateur National d'Ions Lourds (GANIL), CEA/DSM - CNRS/IN2P3,
BP 55027, F-14076 Caen Cedex, France
}%

\date{\today}

\begin{abstract}
Bound states of  dipole-bound negative anions are studied by using a non-adiabatic pseudopotential method and the
Berggren expansion involving bound states, decaying resonant states, and non-resonant scattering continuum. The method is benchmarked by using the traditional technique of direct integration of coupled channel equations.
A good agreement between the two methods has been found for well-bound states. For   weakly-bound subthreshold states with binding energies comparable with rotational energies of the anion, the direct integration approach breaks down
and the Berggren expansion method becomes the tool of choice.
\end{abstract}

\pacs{03.65.Nk,	
31.15.-p, 
31.15.V-,	
33.15.Ry	
}

\maketitle

\section{Introduction}

Weakly-bound many-body systems are intensely studied in different domains of mesoscopic physics \cite{Rii2000,Roti09}, including nuclear  \cite{hans87,tani96,Cob97,jens04,Maz06}, molecular  \cite{Lim77,Moi79,gris00,bres02,li06,Lef09}, and atomic  \cite{Mitr05,Varga08,Ferlaino} physics. In this context, dipolar anions are one of the most spectacular examples of marginally bound quantum systems \cite{garrett70_118,garrett71_111,garrett80_105,Garrett80,Garrett81,garrett82_104,Christ71,Comp77,Wong74,Rohr76,carlsten76_126,Jordan76,Lykke84,brinkman93_208,mullin93_209,desfrancois96_114,Smith97,desfrancois98,ard09_122,Compt01,Jordan03,Chernov09}.  

The mechanism for forming anion states by the long-range dipolar potential has been proposed by Fermi and Teller \cite{fermi47_110}, who studied the capture of negatively charged mesons in matter. They found that if a negative meson is captured by a hydrogen nucleus, the binding energy of the electron becomes zero for the  electric dipole moment of a 
meson-proton system  ${ {\mu}_{cr} = 1.625 }$\,D.  Later this result was generalized to the case of an extended dipole with an infinite moment of inertia \cite{levy67_115}.   Lifting the adiabatic approximation by considering the rotational degrees of freedom of the anion \cite{garrett70_118,garrett71_111,garrett80_105,Garrett80,Garrett81,garrett82_104} turned out to be crucial; it  also boosted  the critical value of ${\mu}$ to  about 2.5 D. 
For anions with ${ \mu > {\mu}_{cr} }$, the number of bound states of the electron becomes finite, and the critical electric dipole moment ${ {\mu}_{cr} }$ depends on the moment of inertia of the molecule. In the  non-adiabatic calculations, the pseudo-potential was used to take into account finite size effects, repulsive core, polarization effects,  and quadrupolar interaction. The pseudo-potential method has provided a convenient description of binding energy of the electron bound by an electric dipolar field. Recently, this method was applied to linear electric quadrupole systems \cite{garrett08_131}.
Some recent theoretical studies of dipole-bound anions also employed the  coupled cluster 
technique \cite{Kalcher00,Kalcher00a,Skurski02}.

The unbound part of the spectrum of multipolar anions has been discussed theoretically in Refs.~\cite{estrada84_294,sadeghpour00_297} and Refs. quoted therein. Resonance energies of dipolar anions have been determined experimentally by low energy electron scattering off the dipolar molecules \cite{Wong74,Rohr76,Lykke84,mullin93_209,Smith97}.

Both the long-range dipole potential and the weak binding of dipolar anions provide a considerable challenge for theory. The impact of the molecular rotation  on a weakly-bound electron can be represented by coupled-channel (CC) equations that can be solved by means of the direct integration. While this approach correctly predicts the number of bound states of polar anions, it is less precise for treatment of  weakly-bound excited states. Moreover, it cannot be used for studies of dipolar anion resonances because the exact asymptotics for a dipolar potential in the presence of a molecular rotor cannot be determined. 

In this paper, we apply the complex-energy configuration interaction framework   based on the Berggren ensemble~\cite{berggren1968}
to the problem of bound states in dipole-bound negative anions. 
The Berggren completeness relation is a resonant-state expansion; it  treats the resonant  and scattering states on the same footing as  bound states. We have successfully applied this tool to a variety of nuclear structure problems pertaining to  weakly-bound and unbound nuclear states 
\cite{Mic02,Mic03,PRC_Lithium_2004,IdB02} (for a recent review see Ref. \cite{Mic09}). The nuclear many-body realisation of the complex energy configuration interaction method is known under the name of the  Gamow Shell Model.

Resonances do not belong to the Hilbert space, so the mathematical apparatus of quantum mechanics in Hilbert space is inadequate for  Gamow states \cite{Gam28}, which are not square-integrable. It turned out that the mathematical structure of the Rigged Hilbert Space (RHS) \cite{gelfand61,maurin68,bohm78} can accommodate time-asymmetric processes, such as particle decays, by
extending the domain of quantum mechanics.   The mathematical setting of the resonant state expansions follows directly from the formulation of quantum mechanics in the RHS \cite{gelfand61,maurin68}, rather than the usual Hilbert space \cite{bohm78,Mad05,Madrid12}.

The Berggren ensemble    provides a natural generalization of the configuration interaction  for the description of the particle continuum. The complex-energy  Gamow-Siegert states \cite{Gam28,Sie39}  states have been used in various contexts in nuclear, atomic, and molecular physics \cite{Pei59,Humblet_Rosenfeld,Gyarmati86,lind1993,Bol96,Mad02,Hamilton02,Kap03,Her03,Kap05,Santra05,Jul07,Mad07,Toyota07}. Some recent applications of Gamow-Siegert states, also  in the context of a CC formalism relevant to the problem of dipole anions, can be found in, e.g.,  Refs. \cite{Kruppa00,Barmore00,Kru04,Tolstikhin06,Tolstikhin08}.

This paper is organized as follows. 
The Hamiltonian of the pseudo-potential method is briefly discussed in Sec.~\ref{S:pseudopotential}. The CC formulation of the 
Schr\"odinger 
equation for dipole-bound negative anions is outlined in Sec.~\ref{S:CC}.
Section~\ref{direct_integration} discusses the direct integration method (DIM)
for solving  the CC problem with a focus on difficulties in  imposing
proper boundary conditions when the rotational motion of the molecule is considered.
The Bergggren expansion method (BEM)  is introduced in Sec.~\ref{Berggren_basis_diag}. Section~\ref{parameters_ps-pot} specifies the  coupling constants of the pseudo-potential and other calculation parameters.
Salient features of DIM and BEM  solutions are compared in Sec.~\ref{TestResults}. The predictions of DIM and BEM  for low-lying energy states  and r.m.s.  radii of LiI$^-$,
LiCl$^-$, LiF$^-$, and LiH$^-$ anions  are collected in Sec.~\ref{S:results}.
Finally, Sec.~\ref{S:conclusions} contains the
 conclusions and outlook.

\section{Hamiltonian}\label{S:pseudopotential}

A dipole-bound negative  anion is composed of a neutral polar molecule 
with a dipole moment greater than ${\mu}_{cr}$  and a valence
electron.
The Hamiltonian of the total system can be written as:
\begin{equation}
	{H}_{tot} = {H}_{e} + {H}_{mol} + {V}
\label{H}
\end{equation}
where ${H}_{e}$ is the Hamiltonian of the valence electron, ${H}_{mol}$ 
is the Hamiltonian of the molecule, and $V$ is the electron-molecule interaction. 
The many-body Schr\"odinger equation for ${H}_{tot}$
couples all electrons of the system; hence, an approximation scheme has to be developed.

As a first simplification, we assume that the
vibrational motion of a  molecule is much slower than
other modes so that it can be treated in the Born-Oppenheimer approximation.
The Hamiltonian (\ref{H}) simplifies considerably  if one considers anions of closed-shell systems. Moreover, 
if spin is neglected \cite{garrett82_104}, the  molecule can be treated as a rigid rotor.
Note that the energy scales associated with the rotational motion of the molecule and the motion of the weakly-bound valence electron
may be comparable. Consequently, 
there appears a strong non-adiabatic coupling between the molecular angular momentum  $\gras{j}$ and the orbital angular momentum $\gras{\ell}$ of the electron.
Eq. (\ref{H}) thus writes within this approximation scheme:
\begin{equation}
	{H_{tot}} = \frac{\gras{p}_e^2}{2m_e} + \frac{\gras{j}^2}{2I} + {V}
\label{H_Born_Opp}
\end{equation}
where ${ I }$ is the moment of inertia of the neutral molecule,
${ \gras{p}_{e} }$ is the linear momentum of the valence electron and ${ {m}_{e} }$ its mass. The interaction
$V$ is approximated by a one-body pseudo-potential ${ V (r , \theta) }$ acting on the valence electron \cite{garrett78,garrett79,garrett82_104}:
\begin{equation}
	V(r , \theta) = {V}_{\mu}(r,\theta) + {V}_{\alpha}(r,\theta) + {V}_{ {Q}_{zz} }(r,\theta) + {V}_{\rm SR}(r),
	\label{eq_main_pot}
\end{equation}
where ${ \theta }$ is the angle between the dipolar charge separation $\gras{s}$
and electron coordinate;
\begin{equation}
	{V}_{\mu}(r,\theta) = -  \mu e  \sum_{\lambda=1,3,\cdots}
	 {\left( \frac{ {r}_{<} }{ {r}_{>} }
			      \right)}^{\lambda} \frac{1}{s {r}_{>}}
	 {P}_{\lambda}( \cos\theta )
\label{Vmu}
\end{equation}
is the dipole potential of the molecule; 
\begin{equation}
	{V}_{\alpha}(r,\theta) = - \frac{ e^2 }{ 2 {r}^{4} } \left[ {\alpha}_{0} + {\alpha}_{2} {P}_{2}( \cos\theta ) \right] f(r)
\label{Valpha}
\end{equation}
is the induced dipole potential, where 
 ${ {\alpha}_{0} }$ and ${ {\alpha}_{2} }$ are the spherical and quadrupole polarizabilities of the linear molecule;
\begin{equation}
	{V}_{ {Q}_{zz} }(r,\theta) = - \frac{e}{ {r}^{3} } {Q}_{zz} {P}_{2}( \cos\theta ) f(r)
\label{VQzz}
\end{equation}
is the potential due to the permanent quadrupole moment of the molecule; and 
a short-range potential
\begin{equation}
	{V}_{\rm SR} (r) = {V}_{0} \exp (- {(r / {r}_{c})}^{6})
\label{VSR}
\end{equation}
accounts for  the exchange effects and
compensates for spurious effects induced by the cut-off function
\begin{equation}
	f (r) = 1 - \exp\{-{(r/r_0)}^6\}
\label{cutoff}
\end{equation}
introduced in Eqs.~(\ref{Valpha},\ref{VQzz})
to avoid a singularity at  ${ r \to 0 }$. The parameter $r_0$ in
Eq.~(\ref{cutoff}) is an effective short-range cutoff distance for
the long-range interactions.

\section{Coupled-channel expression of the Hamiltonian}\label{S:CC}

The eigenfunctions of the Hamiltonian (\ref{H_Born_Opp}) can be conveniently expressed in the CC representation:
\begin{equation}
   {\Psi}^{J} = \sum_{c} u_c^J(r)  \Phi_{j_c \ell_c}^{J} 
\end{equation}
where the index $c$ labels the  channel, $u_c^J(r)$ is the radial wave function of the valence electron in a channel $c$, and the channel function
$\Phi_{j_c \ell_c}^{J}$ arises from the coupling of ${j}_c$ and
${\ell}_c$ to the total angular momentum $J$ of the anion: ${\gras{j}}+{\gras{\ell}}={\gras{J}}$.
Due to rotational invariance of 
${H}_{tot}$,  its matrix elements are
independent of the magnetic quantum number $M$, which will be omitted in the following.

The potential ${ V (r , \theta) }$ in Eqs.~(\ref{eq_main_pot} - \ref{VSR}) 
can be expanded in multipoles:
\begin{equation}
	V(r,\theta) = \sum_{\lambda} V_{\lambda}(r) P_{\lambda}(\cos\theta),
\end{equation}
where 
\begin{equation}
	P_{\lambda}(\cos\theta) = \frac{4\pi}{2\lambda+1} Y_{\lambda}^{(mol)}(\hat{\gras{s}}) \cdot Y_{\lambda}^{(e)}(\hat{\gras{r}}).
\label{Plambda_sph_harmonics}
\end{equation}
The  matrix elements of ${ {P}_{\lambda}(\cos\theta) }$ 
between the  channels $c$ and $c'$ are obtained by means of the standard angular momentum algebra:
\begin{eqnarray}
	&&\braket{ \Phi_{j_{c'} \ell_{c'}}^{J}
		   | P_{\lambda}(\cos\theta) | \Phi_{j_c \ell_c}^{J} }\nonumber \\
		&=& {(-1)}^{j_{c'} + j_c + J}
			\begin{Bmatrix}
				j_{c'} & \ell_{c'} & J \\
				\ell_c & j_c & \lambda
			\end{Bmatrix}
			\begin{pmatrix}
				j_{c'} & \lambda & j_c \\
				0 & 0 & 0
			\end{pmatrix}
			\begin{pmatrix}
				\ell_{c'} & \lambda & \ell_c \\
				0 & 0 & 0
			\end{pmatrix} \nonumber \\
		&\times& \sqrt{ (2\ell_{c'}+1)(2\ell_c+1)(2j_{c'}+1)(2j_c+1) }.
\label{Plambda_ME}
\end{eqnarray}

In the following, we express $r$ in units of the Bohr radius $a_0$,  $I$
in units of $m_e a_0^2$, and  energy in Ry. 
The radial functions  $u_c^J(r)$  are solutions of the set of CC equations:
\begin{eqnarray}
	&&\left[ \frac{d^2}{dr^2} - \frac{
	 {\ell}_{c}({\ell}_{c}+1) }{ {r}^{2} } - \frac{
	 {j}_{c}({j}_{c}+1) }{I} + {E_J} \right] {u}_{c}^J(r) \nonumber \\
 &=& \sum_{c'} v_{cc'}^J (r) u_{c'}^J(r),
\label{CC3}
\end{eqnarray}
where ${ E_J }$ is the energy of the system and
\begin{equation}
	v_{cc'}^J(r) = \sum_\lambda 
	\braket{ \Phi_{j_{c'} \ell_{c'}}^{J}
		   | P_{\lambda}(\cos\theta) | \Phi_{j_c \ell_c}^{J} }
	V_\lambda(r).
\label{Vccp}
\end{equation}

\section{Direct integration of coupled-channel equations} \label{direct_integration}

The CC equations (\ref{CC3}) can be solved by the DI method.
Below we describe the method used to generate the channel wave functions 
$u_c(r)$ (from now on, the quantum number $J$ is omitted to simplify notation) obeying the physical boundary conditions. Namely, we assume that $u_c(r)$ is regular at origin: ${u}_{c} (r = 0) = 0$, and  for ${ r \rightarrow + \infty }$ it behaves like an outgoing wave ${u}_c^+(r)$.

The central issue of DI lies in the boundary condition
at infinity. Indeed, as we shall see in Sec.~(\ref{asymptotic_for_channels}), an asymptotic wave function of a dipole-bound anion is not analytic in general,
so that one cannot exactly impose outgoing boundary conditions. This calls for the use of controlled approximations. 
In the following, we describe the numerical integration of CC equations. While the method is standard (cf. Sec. 3.3.2  of Ref.~\cite{Tho88}), this particular application is not; hence key details should be given.

\subsection{The basis method with the direct integration} \label{basis_functions}

To integrate CC equations, we introduce the matching radius $r_m$  that defines the internal region  $[0:r_m]$, where the centrifugal potential is appreciable,
and the external zone $[r_m:+\infty]$. 
An internal basis function  ${u}^{0}_{b;c} (r)$ in $[0:r_m]$
is regular at $r=0$:
\begin{equation}
	{u}^{0}_{b;c_b} (r) \sim {r}^{{\ell}_{{c}_{b}} + 1}
\end{equation}
in one channel ${c}_{b} = ({j}_{{c}_{b}} , {\ell}_{{c}_{b}})$.
The CC equations imply that when $r \rightarrow 0$ the internal channel wave functions 
${u}^{0}_{b;c} (r)$ with  ${ c \neq {c}_{b} }$ must behave as:
\begin{equation*}
\frac{2m_e V_{c c_b}(0)}{\hbar^2}\times 
\left\{ 
 \begin{array}{l l} 
 \frac{r^{\ell_{c_b} + 3}}{2\ell_{c_b} + 5}  \ln(r/r_m)  & \text{for}~\ell_{c} = \ell_{c_b} + 2,   \\
 \frac{r^{\ell_{c_b} + 3}}{(\ell_{c_b} + 2)(\ell_{c_b} + 3) - \ell_{c}(\ell_{c} + 1)}  & ~\text{otherwise.}  
 \end{array} \right. 
\end{equation*}
Note that is it necessary to pay attention when integrating CC equations close to  ${ r = s }$, as the potential  (\ref{Vmu}) is not differentiable therein.

In the external region $[r_m:+\infty]$, the basis wave functions are denoted
${u}^{+}_{b;c_b} (r)$. By construction, at very large distances of the order of hundreds of $a_0$ (asymptotic region), ${u}^{+}_{b;c_b} (r) \neq 0$ for 
${ {c}_{b} = ({j}_{{c}_{b}} , {\ell}_{{c}_{b}}) }$ and ${u}^{+}_{b;c} (r) = 0$
for other channels ${ c \neq {c}_{b} }$. 
The asymptotic behavior of external channel functions  is discussed  in Sec. \ref{asymptotic_for_channels} below.

Both sets of internal and external  basis functions are used to expand the channel function $u_c(r)$:
\begin{equation}\label{u_expansion} 
u_c(r) = \left\{ 
 \begin{array}{l l}
\sum_b C^0_b u^0_{b;c}(r) & \quad \text{for $r \leq r_m$}, \\ 
\sum_b C^+_b u^+_{b;c}(r) & \quad \text{for $r \geq r_m$}
 \end{array} \right.
\end{equation}
The matching conditions at ${ r = {r}_{m} }$
\begin{eqnarray}\label{other_match_eqs}
\sum_b \left[ C^0_b u^0_{b;c}(r_m) - C^+_b u^+_{b;c}(r_m) \right] &=& 0,  \\
\sum_b \left[ C^0_b \frac{du^0_{b;c}}{dr}(r_m) - C^+_b \frac{du^+_{b;c}}{dr}(r_m) \right] &=& 0, 
\end{eqnarray}
form a linear system of equations: ${ AX = 0 }$. The condition of
${ \det A = 0 }$ determines the energy  of a bound or resonant state.
(One can thus see that $\det A$ is thus the generalization of the Jost function for CC equations.) Once the eigenenergy  has been found, the amplitudes $ {C}^{0}_{b}, {C}^{+}_{b}$ are given by the eigenvector ${ X }$ of ${ A }$. The overall norm is determined by the condition:
\begin{equation}
\sum_c \int_0^{+\infty} |u_c(r)|^2~dr = 1. \label{norm_uc}
\end{equation}

\subsection{The coupled-channel  equations in the asymptotic region} \label{asymptotic_for_channels}

At large distances, ${ {V}_{cc'} (r) }$ can be written as:
\begin{equation}\label{Vccp_asymp}
V_{cc'}(r) = \frac{\hbar^2}{2m_e} \left[ \frac{\chi_{cc'}}{r^2} + V_3(r) \right],
\end{equation}
where $\chi_{cc'}$ is a constant and ${ V_3 (r) }$ decreases for ${ r \rightarrow + \infty }$ as $r^{-3}$.
In the following, we shall  assume that ${ V_3 (r) = 0 }$ in the asymptotic region.
As the numerical integration up to ${ r \sim 100~{a}_{0} }$ is stable, 
the error made by neglecting $V_3$ is around ${ {10}^{-6}~{a}_{0}^{-3} }$, which is sufficiently small to
insure that the asymptotic zone has been practically reached.

Let us first consider the case of  an infinite moment of inertia $I \rightarrow + \infty$. Here, Eq.~(\ref{CC3}) becomes:
\begin{equation}
u''_c(r) = \frac{\ell_c(\ell_c + 1)}{r^2} u_c(r) + \sum_{c'} \frac{\chi_{cc'}}{r^2} u_{c'}(r) - k^2 u_c(r) \label{CC_equations_asymp_app},
\end{equation}
where ${ k = \sqrt{E} }$. 
The outgoing solution of (\ref{CC_equations_asymp_app}) in a basis channel $b$
can be written in terms of  spherical Hankel functions:
\begin{equation}
u^+_{b;c}(r) = g^{(b)}_c H^+_{\ell^{(b)}_{eff}}(kr) \label{ansatz_CC_equations_asymp_app},
\end{equation}
where 
${ {\ell}^{(b)}_{eff} }$ is an effective angular momentum given by eigenvalues of the eigenproblem
\begin{equation}\label{gc_l_eff_eigensystem}
\ell_c(\ell_c+1) g^{(b)}_c + \sum_{c'} \chi_{cc'} g^{(b)}_{c'} = \ell^{(b)}_{eff}(\ell^{(b)}_{eff} + 1) g^{(b)}_c.
\end{equation}
Indeed, it immediately follows from Eqs.~(\ref{CC_equations_asymp_app}) and (\ref{gc_l_eff_eigensystem}) that:
\begin{equation}
u^+_{b;c}(r)'' = \left( \frac{\ell^{(b)}_{eff}(\ell^{(b)}_{eff} + 1)}{r^2} - k^2 \right) u^+_{b;c}(r),
\label{l_eff_eq}
\end{equation}
so the physical interpretation of ${ {\ell}^{(b)}_{eff} }$ in terms of an effective angular momentum is justified.

If ${ I }$ is finite, however, solutions of Eq. (\ref{CC3}) are no longer analytical at large distances.
Nevertheless, it is possible to construct an adiabatic approximation for ${ {u}_{c} (r) }$ in the asymptotic region.
To this end, one defines the linear momentum ${ {k}_{c} = \sqrt{E - {j}_{c} ({j}_{c} + 1)/I} }$ for a channel ${ c }$. In the asymptotic region, Eq.~(\ref{CC3}) becomes:
\begin{equation}
u''_c(r) = \frac{\ell_c(\ell_c + 1)}{r^2} u_c(r) + \sum_{c'} \frac{\chi_{cc'}}{r^2} u_{c'}(r) - k_c^2 u_c(r) \label{CC_equations_asymp},
\end{equation}
where, compared to Eq.~(\ref{CC_equations_asymp_app}), ${ k }$ is replaced by the channel momentum ${ {k}_{c} }$.
This approximation can be applied if ${ |E| \gg {j}_{c} ({j}_{c} + 1)/I }$ for all channels of importance. In those cases, one can introduce an 
ansatz for ${ {u}_{c} (r) }$ by  replacing ${ k }$ by ${ {k}_{c} }$ in Eq.~(\ref{ansatz_CC_equations_asymp_app}). The  relative error 
on a basis function ${ {u}_{+}^{(b;c)} (r) }$ associated with this approximation is
\begin{equation}
\sum_{c'} \left| \frac{\chi_{cc'}}{r^2} \frac{g^{(b)}_{c'}}{g^{(b)}_{c}} 
\left( \frac{H^+_{\ell^{(b)}_{eff}}(k_{c'} r)}{H^+_{\ell^{(b)}_{eff}}(k_c r)} - 1 \right) \right|,
\label{adiabatic_error}
\end{equation}
i.e.,  is of the order of ${ |{k}_{c} - {k}_{c'}|/{r}^{2} }$. 

In practical calculations, ${ I \sim {10}^{5} }$ and ${ {j}_{max} \sim 7 }$. This gives  ${ {j}_{max} ({j}_{max} + 1)/I \sim {10}^{-4} }$. Consequently, if
${ |E| > {10}^{-3} }$ Ry, the error ${ |{k}_{c} - {k}_{c'}|/{r}^{2} < {10}^{-6}~{a}_{0}^{-3} }$ for ${ r \sim 100~{a}_{0} }$ is close to that associated with  the neglect of ${ V_3 (r) }$. On can thus see
that the proposed ansatz accounts for the coupling  term (\ref{Vccp_asymp}) in many cases. However, this approximation  breaks down for  weakly-bound/unbound states with
$|E| < 10^{-4}$ Ry; hence,  a more adequate  theoretical method based on a resonant state expansion needs to be introduced.

\section{Diagonalization with the Berggren basis} \label{Berggren_basis_diag}

Another way to find eigenstates of the CC problem  (\ref{CC3}) is to
diagonalize the associated Hamiltonian in a complete basis of single-particle states. Since our goal is to describe weakly-bound or unbound states, special care should be taken to treat  the asymptotic part of wave functions as
precisely as possible.
A suitable basis for this problem is the one-body Berggren  ensemble  \cite{berggren1968,Beggren1993,lind1993}. 
This basis is generated by a finite-depth spherical potential 
and contains bound ($b$), decaying ($d$), and scattering ($s$) one-body states. Fot that reason, 
the Berggren ensemble is ideally suited to deal with structures having large spatial extensions (such as halos or Rydberg states) or outgoing behavior (such as decaying resonances). Some recent applications, in a many-body context, have been reviewed in Ref.~\cite{Mic09}.

\subsection{The Berggren basis}

The finite-depth potential generating the Berggren ensemble can be chosen arbitrarily.
To improve the convergence, however,  it is convenient in practical applications to use a one-body potential, which is as close as possible to the Hartree-Fock field of the Hamiltonian in question. Therefore, in the case of the one-body problem  (\ref{CC3}), the most optimal  potential to generate the Berggren basis is the
diagonal part of ${ {v}_{cc'} (r) }$. This means that the basis states ${ {\Phi}_{k , c} (r) }$
are eigenstates of the spherical potential ${ {v}_{cc} (r) }$:
\begin{equation}
\Phi_{k,c}''(r) = \left( \frac{\ell_{c}(\ell_{c}+1)}{r^2} + v_{cc}(r) - k^2 \right) \Phi_{k,c}(r)
\label{Berggren_basis_Phi}
\end{equation}
that obey the following boundary conditions:
\begin{eqnarray}
\!\!\!\!\!\!\!\!\! \Phi_{k,c}(r) &\sim& C^0 r^{\ell_c + 1},~r \sim 0 ~~\text{for all states}, \label{condition_in_zero} \\
\!\!\!\!\!\!\!\!\! \Phi_{k,c}(r) &\sim& C^+ H^+_{\ell_c}(k r),~r \rightarrow +\infty \mbox{ (${ b , d }$)}, \label{bound_resonant} \\
\!\!\!\!\!\!\!\!\! \Phi_{k,c}(r) &\sim& C^+ H^+_{\ell_c}(k r)+ C^- H^-_{\ell_c}(k r),~r \rightarrow +\infty \mbox{ (${ s }$)}, \label{scattering}
\end{eqnarray} 
where the boundary conditions at ${ r \sim 0 }$ and at ${ r \rightarrow + \infty }$ for
scattering states (${ s }$) are standard, and for
bound and decaying states (${ b , d }$) 
one imposes the outgoing boundary condition. 
Note that ${ k }$ in Eq.~(\ref{Berggren_basis_Phi}) is in general complex.

The scattering states are normalized to  the Dirac delta, which results in a condition for the $C^-$ and $C^+$ amplitudes in (\ref{scattering}) \cite{berggren1968}:
\begin{equation}\label{Dirac_scat_norm}
\braket{\Phi_{k,c} | \Phi_{k',c}} = \delta(k - k') \Leftrightarrow 2 \pi C^+ C^- = 1.
\end{equation}
The normalization of bound states is standard as well, but that of decaying
resonant states is not. Indeed, resonant states rapidly oscillate and diverge exponentially in modulus
along the real ${ r }$-axis; hence, one cannot calculate their norm in the same way as for the bound states. 

The solution of this problem is provided by the exterior complex scaling
\cite{Gyarmati1971}, i.e., one calculates the norm of the resonant state using complex ${ r }$ radii:
\begin{eqnarray}
\braket{\Phi_{k,c}|\Phi_{k,c}} &=& \int_0^R \Phi^2_{k,c}(r)\,dr \nonumber \\
&+& \int_0^{+\infty} \Phi^2_{k,c}(R + x e^{i\theta})\,e^{i\theta}\,dx,
\label{complex_scaling_norm},
\end{eqnarray}
where ${ R }$ is a radius taken sufficiently large so that condition
(\ref{bound_resonant}) is fulfilled. In the above formula, ${ \theta }$ is an angle of rotation chosen so that ${ {\Phi}_{k , c} (R + x {e}^{i \theta}) \rightarrow 0 }$
for ${ x \rightarrow + \infty }$, which is always possible provided ${ \theta }$ is
larger than a critical value depending on $k$ \cite{PRC_O_He_2003}.
Note that no
modulus enters Eq.~(\ref{complex_scaling_norm}). This arises from the
finite-lifetime character of resonant states, which requires us to use the
biorthogonal scalar product \cite{Agui71,Bals71,Simo72,PRC_O_He_2003,Mic09}.
It can be shown \cite{Agui71,Bals71,Simo72} that the norm defined in  Eq.~(\ref{complex_scaling_norm}) is indeed independent of 
${ R }$ and ${ \theta }$, as expected from a norm.
Since the  expression (\ref{complex_scaling_norm}) is also valid for bound states, 
bound and decaying states enter the Berggren ensemble as one family of resonant states.

The exterior complex scaling can be used to calculate matrix elements of a
one-body operator
${ O (r) }$ as well, provided it decreases faster than ${ 1/r }$ along the complex ${ r }$-contour:
\begin{eqnarray}
&&\braket{\Phi_{k',c'}|O|\Phi_{k,c}} = \int_0^R \Phi_{k',c'}(r)~O_{cc'}(r)~\Phi_{k,c}(r)\,dr \nonumber \\ 
&+& \int_0^{+\infty} \!\!\!\!\! \Phi_{k',c'}(z(x))~O_{cc'}(z(x))~ \Phi_{k,c}(z(x))\,e^{i\theta}\,dx
\label{complex_scaling_operator},
\end{eqnarray}
where ${ z (x) = R + x {e}^{i \theta} }$, and ${ \ket{{\Phi}_{k , c}} }$ and ${ \ket{{\Phi}_{k', c'}} }$ can here be bound, decaying,  or scattering states.

\subsection{The Berggren completeness relation}

%
%
\begin{figure}[htb]
\includegraphics[angle=00,width=0.8\columnwidth]{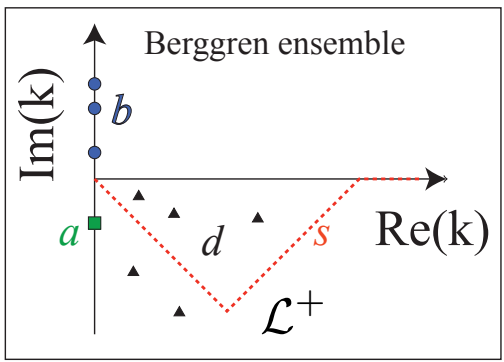}
\caption{Berggren ensemble in the complex momentum plane. The bound ($b$) and antibound ($a$) states are distributed along the imaginary ${ k }$-axis at ${ Im (k) > 0 }$ and ${ Im (k) < 0 }$, respectively. The decaying resonant states ($d$) are located in the fourth quadrant (${ Re (k) > 0 }, { Im (k) < 0 }$. The Berggren completeness relation involves bound states,  scattering states ($s$) on the ${ {\cal L}^{+} }$ contour, and decaying states lying between the real-$k$ axis and ${ {\cal L}^{+} }$. If antibound states are included, the ${ {\cal L}^{+} }$ contour has to be slightly deformed \cite{Mic06,Mic09}. If the contour ${ {\cal L}^{+} }$ lies on the real ${ k }$-axis, the Berggren completeness relation reduces to the Newton completeness relation \cite{Mic09,Newton82}  involving bound and real-energy scattering states.}
\label{Berggren_basis_contour}
\end{figure}
Figure~\ref{Berggren_basis_contour} shows a distribution the Berggren ensemble
in the complex momentum plane. To determine the basis, one 
first  chooses a ${ {\cal L}^{+} }$  contour in the fourth quadrant containing the decaying eigenstates. The scattering states of the ensemble lie on this contour. The resonant part of the ensemble contains  the bound states lying on the imaginary-$k$ axis and  those decaying  states of (\ref{Berggren_basis_Phi})
that are found in the region between the real-${ k }$ axis and  ${ {\cal L}^{+} }$. The Berggren basis is built from all those states:
\begin{equation}
\sum_{n \in (b,d)} \ket{\Phi_{k_n,c}} \bra{\Phi_{k_n,c}} + \int_{{\cal L}^+} \ket{\Phi_{k,c}} \bra{\Phi_{k,c}}~dk = 1.
\label{Berggren_completeness}
\end{equation}
This completeness relation corresponds to a given 
channel ${ c }$; hence, one has to construct Berggren ensembles for all the
 channels considered in Eq.~(\ref{CC3}). 

In order to be able to use  (\ref{Berggren_completeness}) in practice,
one needs to discretize  ${ {\cal L}^{+} }$. Our method of choice is to apply  
the  Gauss-Legendre quadrature to each of the
segments defining   ${ {\cal L}^{+} }$  in Fig.~\ref{Berggren_basis_contour}.
The last segment, chosen along the real-$k$ axis, extends to the large cutoff momentum 
${ k = {k}_{max} }$ that is  sufficiently large
 to guarantee completeness to desired  precision. It is then
convenient to renormalize scattering states using the corresponding  Gauss-Legendre weights ${ {\omega}_{{k}_{n}} }$:
\begin{equation}
\ket{\Phi_{n,c}} = \sqrt{\omega_{k_n}} \ket{\Phi_{k_n,c}}
\label{scat_renorm}.
\end{equation}
The discretized Berggren completeness relation, used in practical computations,  reads:
\begin{equation}
\sum_{i=1}^N \ket{\Phi_{i,c}} \bra{\Phi_{i,c}} \simeq 1
\label{discretized_Berggren_completeness},
\end{equation}
where the ${ N }$ basis states ${ \ket{{\Phi}_{i , c}} }$ 
include all bound, decaying,  and discretized scattering states of the
channel ${ c }$. By using Eq.~(\ref{scat_renorm}), the Dirac delta normalization
of scattering states has been replaced by the usual normalization
to Kronecker's delta. in this way, 
 all ${ \ket{{\Phi}_{i , c}} }$ states can be treated on the same footing
in Eq.~(\ref{discretized_Berggren_completeness}), as in any basis of discrete states.

\subsection{Hamiltonian matrix in the Berggren basis}

As the basis states ${ \ket{{\Phi}_{i , c}} }$ are generated by ${ {v}_{cc} (r) }$, the Hamiltonian matrix  within the  same channel ${ c }$ is diagonal:
\begin{equation}
 \braket{\Phi_{i',c}|h|\Phi_{i,c}}
= \left( k_i^2 + \frac{j_c(j_c+1)}{I} \right) \delta_{ii'} \label{diagonal_ME}
\end{equation}
Matrix elements between two basis states belonging to different channels ${ c }$ and ${ c' }$ are:
\begin{eqnarray}
&& \braket{\Phi_{i',c'}|h|\Phi_{i,c}} = \braket{\Phi_{i',c'}|v|\Phi_{i,c}} \nonumber \\ 
&=& \int_0^R \Phi_{i',c'}(r)~v_{cc'}(r)\,\Phi_{i,c}(r)\,dr \nonumber \\ 
&+& \int_0^{+\infty}  \!\!\!\!\! \Phi_{i',c'}(z(x))~v_{cc'}(z(x))\,\Phi_{i,c}(z(x))\,e^{i\theta}\,dx,
\label{off_diagonal_ME}
\end{eqnarray}
where the complex scaling (\ref{complex_scaling_operator}) can be used, because ${ {v}_{cc'} (r) }$ decreases at least as fast as ${{r}^{-2} }$.

As the off-diagonal matrix elements are present only  for ${ c \neq c' }$, the Berggren basis generated by Eq.~(\ref{Berggren_basis_Phi}) is optimal.
The channel wave functions ${ {u}_{c} (r) }$  can be expressed in the Berggren basis  by diagonalizing
the matrix of ${ h }$ (\ref{diagonal_ME},\ref{off_diagonal_ME}).

\section{Calculation parameters}
\label{parameters_ps-pot}

Results of the direct integration method (DIM)  depend both on the parameters of the pseudo-potential (\ref{eq_main_pot}) and on the cutoff value of the electron orbital angular momentum $\ell_{max}$ considered in the CC problem. They are fixed to reproduce the experimental value of the ground state energy of the LiCl$^-$ anion:
$E_{\rm exp} = -4.483\cdot 10^{-2}$\,Ry \cite{carlsten76_126}. 

The most important term in (\ref{eq_main_pot}) is the dipole potential ${ {V}_{\mu} }$, which depends only on the dipole moment ${ \mu }$ and the size ${ s }$ of the neutral molecule. The remaining parameters of the pseudo-potential
are taken from Ref.~\cite{garrett82_104}, namely:
${\alpha}_{0}=15.3\,{a}_{0}^{3}$, ${\alpha}_{2}=1.1 \,{a}_{0}^{3}$, ${r}_{0}=2.2 \,{a}_{0}$, ${r}_{c}=2.828 \,{a}_{0}$,  ${Q}_{zz}=3.28 \,e {a}_{0}^{2}$, and ${V}_{0}=2.0$\,Ry. The moment of inertia parameters are: $I=150,000 \,m_e a_0^2$ for LiCl$^-$,
$240,000\,m_e a_0^2$ for LiI$^-$, $82,000\,m_e a_0^2$ for LiF$^-$, and $26,000\,m_e a_0^2$ for LiH$^-$.
The dipole moment of each  molecule considered in this work is known experimentally and has been taken from the NIST database. 
%
\begin{figure}[htb]
\includegraphics[angle=00,width=0.8\columnwidth]{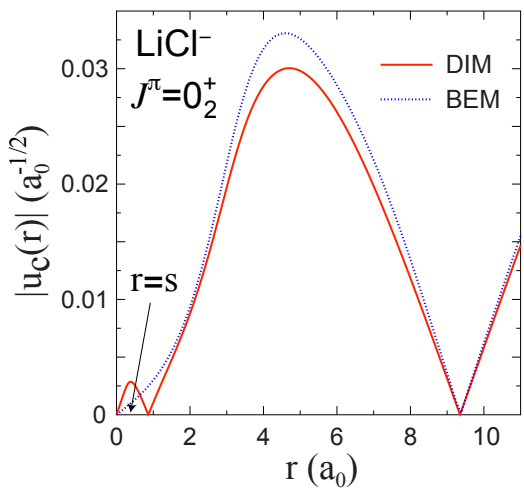}
\caption{The modulus of the channel wave function ${u}_{j=0,\ell=0}$ near $r=0$ for the first excited $J^{\pi} = 0_2^+$ state of LiCl$^-$   calculated in DIM (solid line) and BEM (dotted line)  with  ${ {\ell}_{max} = 9 }$. The charge separation  ${ s }$ of LiCl has been adjusted in both approaches to the experimental ground state energy in the limit ${ {\ell}_{max} \rightarrow \infty }$.
}
\label{non_analytic_difference}
\end{figure}

For $\ell_{max}=9$ the ground state energy of the LiCl$^-$ anion is reproduced by taking  the charge separation $s_{\rm DIM}^{(9)}=0.336\,a_0$.
To remove the dependence of results on $\ell_{max}$ in the DIM, the ground state energy of LiCl$^-$  is extrapolated for $\ell_{max}\rightarrow\infty$, and the size of the charge separation $s$  is adjusted to reproduce the experimental binding energy. In this case,  $s_{\rm DIM}^{(\infty)}=0.337\,a_0$.
The matching radius was taken as $r_m =  \,a_0$. This value was found to optimize the DIM procedure.

Anion spectra in the BEM depend sensitively on the cutoff parameter $k_{max}$ of the single-particle basis. However, as we shall see in Sec.~\ref{TestResults}, for a chosen value of $k_{max}$ they are practically independent of $\ell_{max}$. In this study, we have chosen $k_{max}=1.53\, a_0^{-1}$ for each partial wave in order to attain both a good numerical precision and approximately the same value of the dipole size parameter $s$ as in DI. In this case, $s_{\rm BEM}^{(9)}=s_{\rm BEM}^{(\infty)}=0.336 \,a_0$. We have used complex contours with straight segments connecting points: $k_1=(0,0)$, $k_2=(0.15,-i0.04)$, $k_3=(1,0)$, and $k_4=k_{max}$ in units of $a_0^{-1}$.
Each scattering contour has been discretized with 220 points. The precise form of the contour does not change results; since the applications carried out in this work pertain to bound states only, we could have used real scattering contours, i.e., the Newton completeness relation \cite{Mic09,Newton82}.

\section{Numerical tests and benchmarking}\label{TestResults}

Along with the asymptotic behavior of channel wave functions, treated
approximately with the DIM  and exactly
within BEM,
the Hamiltonian  (\ref{H_Born_Opp}) cannot be identically
represented in both approaches. Indeed, since the potential ${ {V}_{\mu} (r) }$  (\ref{Vmu}) is not differentiable at ${ r = s }$, it cannot be treated exactly in BEM because the channel wave functions expanded in the Berggren basis are analytic by construction.
In practice, this translates into a node  beyond  ${ r = s }$ in DIM channel wave functions, which is absent in BEM. 
\begin{figure}[htb]
\includegraphics[angle=00,width=0.8\columnwidth]{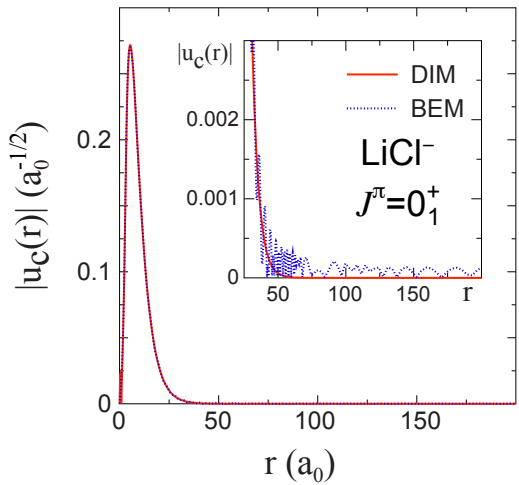}
\caption{
The modulus of the channel wave function ${u}_{j=0,\ell=0}$ for the $J^{\pi} = 0_1^+$ ground state  of LiCl$^-$   calculated in DIM (solid line) and BEM (dotted line)  with  ${ {\ell}_{max} = 9 }$.
At large distances,  spurious wiggles appear in BEM results (see the inset) due to basis truncation. }
\label{Berggren_diag_wiggles}
\end{figure}
This is illustrated in
Fig.  \ref{non_analytic_difference} for a  $(j = 0 , \ell = 0)$ channel function corresponding to the first excited $J^{\pi} = 0_2^+$ state of LiCl$^-$. 
It is to be noted, however, that beyond this point the  channel wave functions calculated with both methods are very close  and -- as will be discussed later -- this near-origin pathology has a very small impact on the total energy as the contribution from this region is small.

As discussed in Sec. \ref{asymptotic_for_channels}, DIM is inadequate for states with very small
energies, while BEM has been shown to be
very precise in this case. On the other hand, for states with binding energies
typically greater than $10^{-2}$ Ry, BEM 
yields channel wave functions that exhibit  spurious
low-amplitude oscillations. Figure \ref{Berggren_diag_wiggles} illustrates such wiggles in the tail of  the channel wave function ${u}_{j=0,\ell=0}$ of the $J^{\pi} = 0_1^+$ ground state  of LiCl$^-$. 
For such  well-bound states 
that  quickly decay with $r$, 
the standard size of the 
Berggren basis (measured in terms of contour discretization points and 
$k_{max}$) is not sufficient.
The DIM is thus preferable for such cases, as 
the asymptotic behavior of well-bound states  is treated almost exactly (see Sec. \ref{asymptotic_for_channels}).

\begin{figure}[htb]
\includegraphics[angle=00,width=0.8\columnwidth]{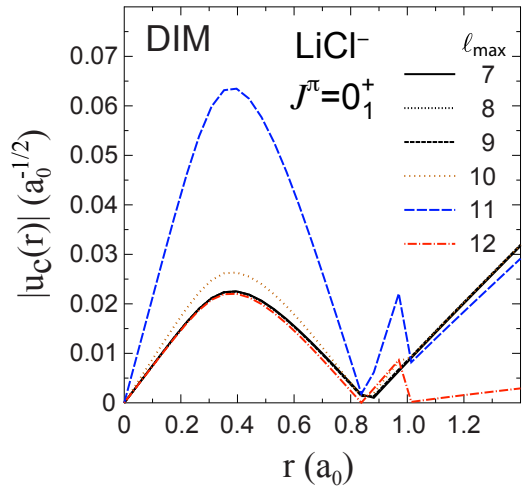}
\caption{
The modulus of the channel wave function ${u}_{j=0,\ell=0}$ for the $J^{\pi} = 0_1^+$ ground state  of LiCl$^-$   calculated in DIM
with several values of ${ {\ell}_{max} }$. For ${ {\ell}_{max} \geq 10 }$, one may notice the development of a discontinuity at the matching point  $r_m= a_0$. In such cases, the channel wave function  becomes ill-conditioned, introducing serious errors in CC eigenenergy and  eigenfunction. }
\label{matching_point}
\end{figure}

The direct integration becomes numerically unstable when the channel orbital angular momentum becomes large, around  ${\ell}_{c} = 10$, 
even for the states with relatively large binding energies. In this case, the 
matrix of basis channel wave functions ${ {u}^{0}_{b;c} (r_m) }$ and ${ {u}^{+}_{b;c} (r_m) }$ and their derivatives, introduced in Sec. \ref{basis_functions} in the context of matching conditions at $r=r_m$,
is ill-conditioned and its eigenvector of zero eigenvalue
becomes imprecise. This results in a discontinuity
at  ${r}_{m}$ and spurious occupation of
channels with large orbital angular momentum ${\ell}_{c} > 10$. This is illustrated in Fig.~\ref{matching_point} for the $J^{\pi} = 0_1^+$ ground state  of LiCl$^-$. A a result, the energy and spatial extension of the electron cloud distribution of the  CC eigenstate become incorrect.

\begin{figure}[htb]
\includegraphics[angle=00,width=0.8\columnwidth]{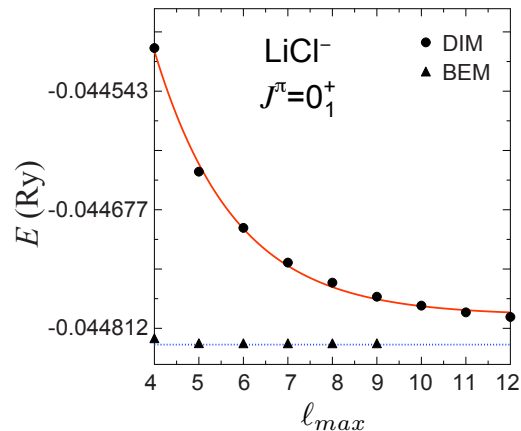}
\caption{The dependence of the LiCl$^-$ ground state energy on $\ell_{max}$ in DIM (dots) and BEM (triangles). The DIM results converge exponentially (red line). This allows us to determine the asymptotic value of energy at $\ell_{max}\rightarrow \infty$. }
\label{l_convergence}
\end{figure}

The convergence of the LiCl$^-$ ground state energy with respect to $\ell_{max}$ is shown in Fig. \ref{l_convergence}.
One may notice an exponential convergence of calculated DIM energies with  $\ell_{max}$ for $6\leq\ell_{max}\leq10$ and a clear deviation for $\ell\geq 11$, which is related to the discontinuity of channel wave functions for $\ell_c>10$.
The energy calculated in BEM is perfectly stable with $\ell_{max}$. 

The rapid converge of BEM with $\ell$  is due to  $k_{max}$-truncation of the single-particle basis that suppresses contributions from  large-$\ell$ configurations. This is illustrated  in Fig. \ref{off-diag_cc-coupling}, which 
displays the average modulus of the off-diagonal matrix element of the channel-channel coupling in BEM:
\begin{figure}[htb]
\includegraphics[angle=00,width=0.8\columnwidth]{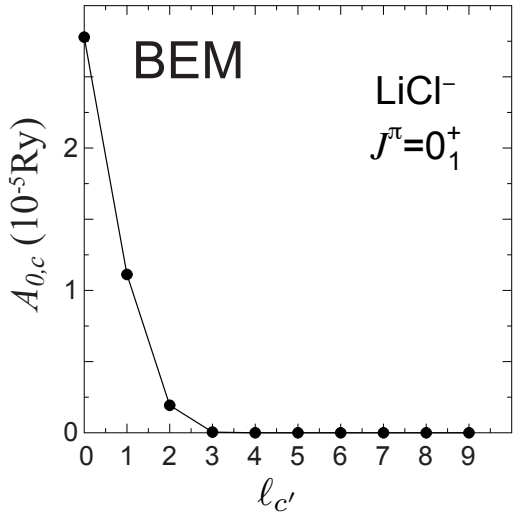}
\caption{Average off-diagonal matrix element $A_{0,c}$ (\ref{ME_average})
of the channel-channel coupling in BEM  between the channel $(j=0,\ell=0)$ and $c'$ for the $J^{\pi}=0_1^+$ ground state of LiCl$^-$.}
\label{off-diag_cc-coupling}
\end{figure}
\begin{equation}
A_{c,c'}=\frac{1}{N^2}\sum_{n,n'}^N|\langle\Phi_{n',c'}|V|\Phi_{n,c}\rangle|
\label{ME_average}
\end{equation}
between the first channel $c=(j=0,\ell=0)$ and higher-$\ell$ channels $c'$.
Only the channels with $\ell_c\leq5$ and $|\ell_c-\ell_{c'}|\leq3$ contribute significantly to the channel coupling matrix element. We checked that this is generally the case. Using the same truncation, the DIM  yields numerically stable results. In this case, the energies of well-bound states ($|E|>10^{-2}$ \,Ry) agree in both methods.

\begin{figure}[htb]
\includegraphics[angle=00,width=0.7\columnwidth]{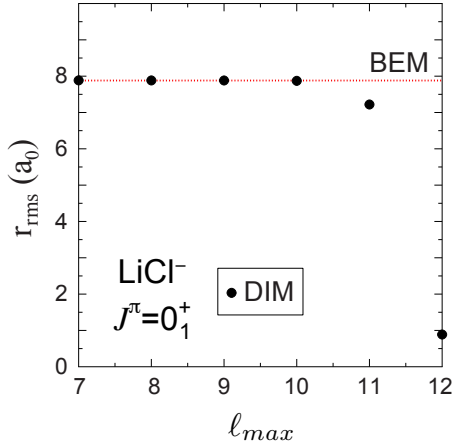}
\caption{The dependence of the LiCl$^-$ ground state r.m.s. radius on $\ell_{max}$ DIM (dots) and BEM (dotted line). The DIM results are stable up to $\ell_{max}=10$. }
\label{radii}
\end{figure}

The numerical instability of DIM at large $\ell_{max}$ leads to a collapse of calculated radii. Figure~\ref{radii} shows the dependence of the ground state r.m.s. radius of  LiCl$^-$  on  $\ell_{max}$. 
This result, together with discussion of Fig.~\ref{off-diag_cc-coupling}, suggests
that the BEM can provide practical guidance on the minimal number of channels in the CC approach.

\begin{figure}[htb]
\includegraphics[angle=00,width=\columnwidth]{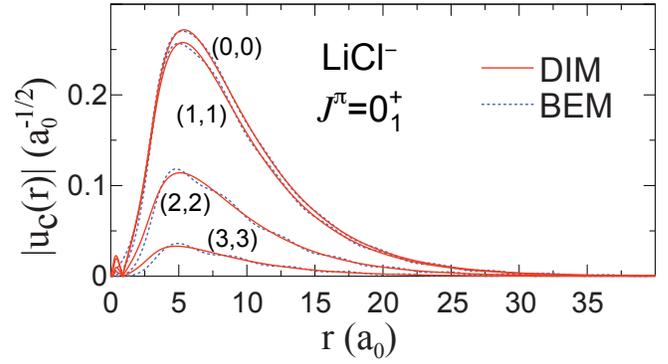}
\caption{Most important channel wave functions $u_c(r)$ 
with $c=(j,\ell)$
for the $J^{\pi}=0_1^+$ ground state of LiCl$^-$, as calculated in DIM (solid line) and BEM (dashed line) with $\ell_{max}=9$. }
\label{LiCl_ground_state}
\end{figure}
In practical applications, spurious oscillations in BEM channel wave functions for well-bound states can be taken care of  by extrapolating  wave functions from the intermediate region of $r$, where they are reliably calculated, into the asymptotic region. This can be done by applying the analytical expression:
\begin{equation}
{\tilde u}_c(r)\equiv \lim_{r\gg 0}u_c(r)=e^{ik_cr}\sum_{j=1}^M\frac{\alpha_j^{(c)}}{r^j},
\label{fit}
\end{equation}
where
$k_c$ is the channel  momentum and $\alpha_j^{(c)}$ are parameters to be determined by the  fit. The precision of this procedure can be assessed by computing the norm of the eigenstate. Using this procedure, one obtains  perfectly stable r.m.s. radii in BEM for different values of $\ell_{max}$, as can be seen in Fig. \ref{radii}.

\begin{figure}[htb]
\includegraphics[angle=00,width=0.9\columnwidth]{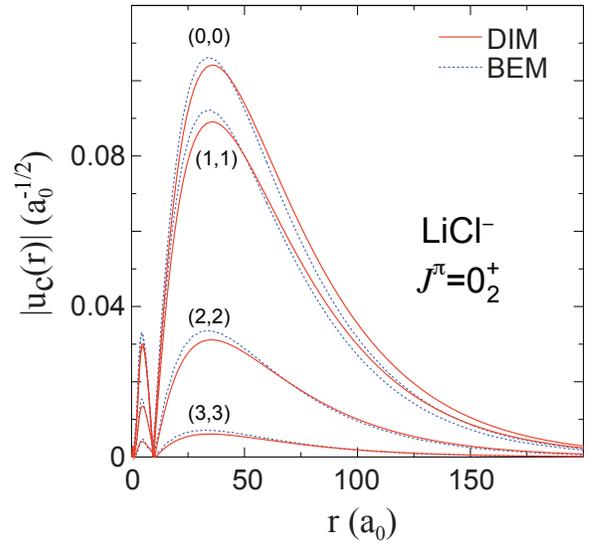}
\caption{Similar to Fig.~\ref{LiCl_ground_state} but for the first excited $J^{\pi}=0_2^+$ state of LiCl$^-$.}
\label{LiCl_first_excited_state}
\end{figure}
\begin{figure}[htb]
\includegraphics[angle=00,width=0.9\columnwidth]{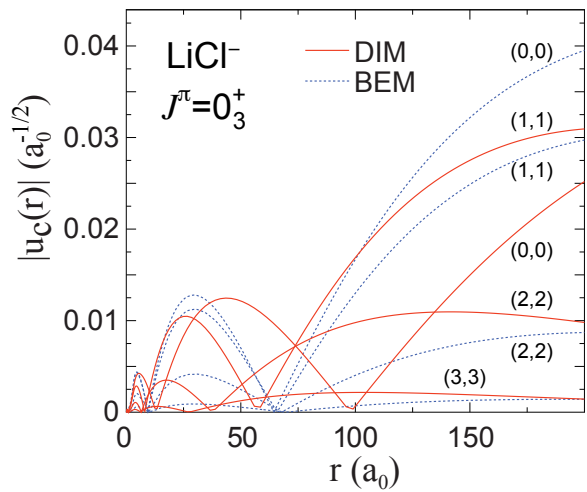}
\caption{Similar to Fig.~\ref{LiCl_ground_state} but for the second excited $J^{\pi}=0_3^+$ state of LiCl$^-$.}
\label{LiCl_second_excited_state}
\end{figure}
Figures~\ref{LiCl_ground_state}-\ref{LiCl_second_excited_state} compare the four most important channel wave functions $(\ell,j)$ of DIM and BEM corresponding to the three lowest $J_i^{\pi}=0_i^+$ eigenstates of LiCl$^-$. For the ground state, 
both approaches predict the same energy $E=-4.483\cdot 10^{-2}$\,Ry and
the channel functions are practically identical.
For the first excited state, the agreement is still reasonable. Here, the energy
in DIM is $E=-7.374\cdot 10^{-4}$\,Ry while BEM gives slightly more binding:
$E=-8.241\cdot 10^{-4}$\,Ry. Consequently, the BEM wave functions decay faster than those computed with DIM. 
For a second excited $0_3^+$ state, both methods differ markedly.
This state has a sub-threshold nature, with $E_{\rm DIM}=-7.051\cdot 10^{-6}$\,Ry and $E_{\rm BEM}=-9.907\cdot 10^{-6}$\,Ry.
For this extremely diffused  state,
the direct integration method fails completely. This is manifested by the very different nodal structure of channel wave functions in DIM seen in Fig.~\ref{LiCl_second_excited_state}.

A stringent test of the computational framework to describe dipolar molecules  is provided by the  analytic result  ${\mu}_{cr}=0.639\,ea_0$ for the fixed dipole  ($I\rightarrow \infty$) \cite{levy67_115}.
To this end, we performed BEM  calculations for a dipolar system at steadily 
decreasing moments of inertia
\cite{garrett71_111,garrett80_105}. 
For each value of $I$, the dipolar anion energies have been calculated for 1080 values of $\mu$ in the interval $0.6\leq\mu\leq3.0$. Only 170 energies satisfying the subthreshold condition $E<E_{lim}=-10^{-8}$\,Ry were retained to minimize the numerical error. These energies correspond to an interval $\Delta\mu\simeq 0.377$ of the dipole moment. 
We checked, that in this energy interval,  $\mu_{cr}$ can be  obtained by 
using the expression
\begin{equation}
	E(\mu) = {(\mu + b)}^{\frac{a}{\mu}} {e}^{c}
	\label{fit_funct}
\end{equation}
to extrapolate the calculated energy down to $E=0$. 
One should stress however, that an excellent energy fit 
in the subthreshold region does not guarantee an excellent estimate of the critical dipole moment. 
The values of $\mu_{cr}$ extracted by this extrapolation procedure can be considered reliable only if $\Delta\mu$, which depends on the chosen precision $E_{lim}$, is close to the critical dipole moment. In the cases studied, this criterion is approximately satisfied only for the ground state and the first excited $0^+$ state. 
The critical dipole moments for these states in  anions with the dipole length $s=4a_0$ are shown in Table \ref{mu_cr} for various moments of inertia. 
\begin{table}[htb]
 \caption{\label{mu_cr}
Critical dipole moments for dipolar anions in the two lowest  $0^+$ 
states calculated in this work (BEM) and in Ref.~\cite{garrett80_105} for the charge separation $s=4a_0$ and different moments of inertia $I$. 
The analytic result  at $I\rightarrow \infty$  \cite{fermi47_110,levy67_115}
is ${\mu}_{cr}=0.639\,ea_0$.
}
\begin{ruledtabular}
\begin{tabular}{c|cc|cc}
\multirow{2}{*}{$I(m_ea_0^2)$} &  \multicolumn{2}{c|}{ ${\mu}_{cr}^{(0)} \, (e {a}_{0})$} &   \multicolumn{2}{c} { ${\mu}_{cr}^{(1)} \, (e {a}_{0})$} \\
&	BEM & Ref. \cite{garrett80_105}   &  BEM & Ref. \cite{garrett80_105}   \\ 
		\hline \\[-6pt]
   ${ {10}^{4} }$ & 0.937 & 0.843 & 1.024 & 1.515 \\
   ${ {10}^{6} }$	& 0.674 & 0.750 & 0.633 & 1.145 \\
   ${ {10}^{8} }$ & 0.639 & 0.715 & 0.622 & 0.974 \\
   ${ {10}^{10} }$  & 0.639 & --- & 0.622 & --- \\
   ${ {10}^{15} }$  & 0.639 & --- & 0.62 & ---  
\end{tabular}
\end{ruledtabular}
\end{table}		
The agreement with the analytic limit is excellent for the ground state configuration, and is fairly good for the first excited 0$^+$ state. This is very encouraging, considering the slow convergence with $I$ and various sources of numerical errors in the $E\rightarrow 0$ regime~\cite{garrett71_111}.

\section{Results for spectra and radii of dipolar anions}\label{S:results}

Energies and r.m.s. radii of the lowest bound $0^+$ and $1^-$ states of LiI$^-$, LiCl$^-$, LiF$^-$, and LiH$^-$ dipolar anions predicted in this study are listed in Table~\ref{tab_spectra}. 
\begin{table}[htb]
 \caption{\label{tab_spectra}
Energies and r.m.s. radii for $0^+$ and $1^-$ bound states of selected dipolar anions obtained in DIM ($\ell_{max}=9$) and BEM. The parameters of the calculation are given in Sec.  \ref{parameters_ps-pot}. The numbers in parentheses denote powers of 10.}
 \begin{ruledtabular}
\begin{tabular}{cc|cc|cc}
\multirow{2}{*}{Anion} & \multirow{2}{*}{state} &  \multicolumn{2}{c|}{$E$\,(Ry)} &   \multicolumn{2}{c} {$r_{\rm rms}\,({a}_{0})$} \\
&	&  DIM & BEM   &  DIM & BEM   \\ 
		\hline \\[-6pt]
   LiI$^-$  &  $0_1^+$ & -5.079(-2) & -5.023(-2) & 7.569(0) & 7.620(0) \\
		    &  $0_2^+$ & -9.374(-4) & -1.037(-3) & 5.112(1) & 4.759(1) \\
		    &  $0_3^+$ & -1.502(-5) & -1.797(-5) & 3.719(2) & 3.308(2) \\
            &  $1_1^-$ & -5.079(-2) & -4.995(-2) & 7.569(0) & 7.641(0) \\
		    &  $1_2^-$ & -9.291(-4) & -1.023(-3) & 5.112(1) & 4.886(1) \\
		    &  $1_3^-$ & -1.261(-7) & -1.099(-5) & 3.423(3) & 3.464(2) \\[2pt]		    
 LiCl$^-$   &  $0_1^+$ & -4.483(-2) & -4.483(-2) & 7.885(0) & 7.894(0) \\
	    	&  $0_2^+$ & -7.374(-4) & -8.241(-4) & 5.632(1) & 5.017(1) \\
		    &  $0_3^+$ & -7.051(-6) & -9.907(-6) & 5.124(2) & 4.106(2) \\
		    &  $1_1^-$ & -4.482(-2) & -4.458(-2) & 7.885(0) & 7.915(0) \\
		    &  $1_2^-$ & -7.241(-4) & -8.067(-4) & 5.633(1) & 5.337(1) \\
		    &  $1_3^-$ & -3.062(-7) & -8.159(-7) & 2.066(3) & 8.831(2) \\[2pt]		    		    
    LiF$^-$ &  $0_1^+$ & -2.795(-2) & -2.983(-2) & 9.117(0) & 8.991(0) \\
		    &  $0_2^+$ & -3.022(-4) & -3.525(-4) & 8.098(1) & 7.501(1) \\
		    &  $0_3^+$ & ---       & -6.101(-8) &  ---    & 3.363(3) \\    
		    &  $1_1^-$ & -2.793(-2) & -2.968(-2) & 9.117(0) & 9.010(0) \\
		    &  $1_2^-$ & -2.782(-4) & -3.277(-4) & 8.124(1) & 7.520(1) \\[2pt]		    		    
	LiH$^-$ &  $0_1^+$ & -2.149(-2) & -2.370(-2) & 1.011(1) & 9.698(0) \\
		    &  $0_2^+$ & -1.491(-4) & -1.922(-4) & 1.058(2) & 9.297(1) \\
  		    &  $1_1^-$ & -2.142(-2) & -2.353(-2) & 1.011(1) & 9.717(0) \\
		    &  $1_2^-$ & -7.942(-5) & -1.231(-4) & 1.146(2) & 9.591(1) 
\end{tabular}
\end{ruledtabular}
\end{table}		

One can see that for each total angular momentum $J^{\pi}$ there are at most three bound eigenstates in each system. The r.m.s. radius of an electron cloud shows a spectacular increase with decreasing the binding energy of the state. For the subthreshold states, such as  $0_3^+$ and $1_3^-$, the radius is of the order of hundreds to thousands $a_0$.

Energy spectra and radii of dipolar anions do not change significantly in the limit $\ell_{max}\rightarrow\infty$. Usually, the extrapolated results for both $E$ and $r_{rms}$ agree very well with those in Table~\ref{tab_spectra} ($\ell_{max} = 9$). For instance, the extrapolated values for the $1_2^+$ state in  LiH$^-$ are  $E=-7.931\cdot 10^{-5}$\,Ry and $r_{\rm rms}=1.147\cdot 10^2\,a_0$.  

The DIM and BEM  results  are generally consistent for both energy and radii though significant quantitative differences  persist for excited, weakly-bound states of anions where the DIM  is not expected to work. In the case of LiF$^-$, the BEM predicts the existence of the third $0_3^+$ state at an energy 
$-6.1\cdot 10^{-8}$\, Ry, which is absent in  DIM.

It is instructive to compare our DIM results with those found in 
Ref.~\cite{garrett82_104} using a similar approach. Table \ref{tab_spectra1} 
lists energies of the lowest $0^+$ bound states of LiI$^-$, LiCl$^-$, LiF$^-$, and LiH$^-$ dipolar anions obtained in both studies, and Table \ref{tab_spectra2} shows the adopted values of dipole moments.

\begin{table}[htb]
 \caption{\label{tab_spectra1}
Energies for $0^+$ bound states of selected dipolar anions obtained in DIM in this work ($\ell_{max}=9$) and in Ref.~\cite{garrett82_104}. The numbers in parentheses denote powers of 10.}
 \begin{ruledtabular}
\begin{tabular}{cc|cc}
\multirow{2}{*}{Anion} & \multirow{2}{*}{state} &  \multicolumn{2}{c}{$E$\,(Ry)}  \\
&	&  This work & Ref. \cite{garrett82_104}      \\ 
		\hline  \\[-6pt]
   LiI$^-$  &  $0_1^+$ & -5.079(-2) & -4.998(-2) \\
		    &  $0_2^+$ & -9.374(-4) & -1.022(-3) \\
		    &  $0_3^+$ & -1.502(-5) & -1.999(-5) \\[2pt]
   LiCl$^-$   &  $0_1^+$ & -4.483(-2) & -4.483(-2) \\
	    	&  $0_2^+$ & -7.374(-4) & -7.497(-4) \\
		    &  $0_3^+$ & -7.051(-6) & -9.775(-6) \\[2pt]
    LiF$^-$ &  $0_1^+$ & -2.795(-2) & -2.793(-2) \\
		    &  $0_2^+$ & -3.022(-4) & -3.366(-4) \\
		    &  $0_3^+$ & ---       & -8.746(-7) \\ [2pt]       
LiH$^-$ &  $0_1^+$ & -2.149(-2) & -2.352(-2) \\
		    &  $0_2^+$ & -1.491(-4) & -1.926(-4) \\
  		    \end{tabular}
\end{ruledtabular}
\end{table}		
 
The two calculations agree reasonably well for the lowest-lying states; some difference stems  from slightly different dipole moments used in  Ref.~\cite{garrett82_104} and here.
Indeed, while the charge separation in both studies was adjusted to reproduce the experimental ground state energy of LiCl$^-$, the fitted values of $s$ in both calculations are different: $s=0.3335~a_0$ in Ref.~\cite{garrett82_104}
and $s_{\rm DIM}=0.336~a_0$ here.  

The largest deviations, seen for weakly-bound states, can be traced back to
the cutoff value of the electron orbital angular momentum when solving CC equations. In 
Ref. \cite{garrett82_104}, adopted
$\ell_{max}$ was small, typically $\ell_{max}=4$ \cite{ard09_122}, whereas it is  fairly large, $\ell_{max}=9$, in our work. As seen in Fig.~\ref{l_convergence} and discussed in Sec. \ref{TestResults},  energies of weakly-bound states obtained in DIM do converge slowly with $\ell_{max}$. Therefore, calculations employing low $\ell_{max}$ values cannot be useful when performing extrapolation
$\ell_{max}\rightarrow\infty$.

\begin{table}[htb]
 \caption{\label{tab_spectra2}
Dipole moments of selected dipolar anions adopted in this work and  in Ref. \cite{garrett82_104}. }
\begin{ruledtabular}
\begin{tabular}{c|cc}
\multirow{2}{*}{Anion} &  \multicolumn{2}{c}{$\mu$\,($ea_0$)}  \\
&	This work & Ref. \cite{garrett82_104}      \\ 
		\hline  \\[-6pt]
		LiI$^-$ & 2.911384272 & 2.911384272 \\
		LiCl$^-$ & 2.805158089 & 2.793355179 \\
		LiF$^-$ & 2.472316049 & 2.478610934 \\
		LiH$^-$ & 2.313370205 & 2.321238811 
	\end{tabular}
\end{ruledtabular}
\end{table}

\section{Conclusions}\label{S:conclusions}

In this study, we applied the theoretical open-system framework based on the Berggren ensemble to a problem of weakly-bound states of dipole-bound negative anions. The method has been benchmarked by using the traditional technique of direct integration of CC equations. While a fairly good agreement between the two methods has been found for well-bound states, the direct integration technique breaks down 
for weakly-bound states with energies $|E| < 10^{-4}$\,Ry, which is comparable with the rotational energy of the anion. For those subthreshold configurations, the Berggren expansion is an obvious tool of choice.

The inherent problem of the DIM is the lack of stability of results when
the  number of channels is increased. Indeed, the method
breaks down  when the channel orbital angular momentum becomes large; around $\ell_c = 10$. This can be traced back to the applied matching condition. We demonstrated that this pathology is absent in BEM. Here, the rapid converge with $\ell$ is guaranteed by an effective softening of the interaction through the momentum cutoff $k_{max}$, which suppresses contributions from high-$\ell$ partial waves. 

The future applications of BET will include the structure of quadrupole-bound
anions \cite{ferron2004,Desfrancois04,garrett08_131,garrett12} and the continuum structure of  anions, including the characterization of low-lying  resonances.

\begin{acknowledgments}
Stimulating discussions with and helpful suggestions from R.N. Compton
and W.R. Garrett, who encouraged us to apply the complex-energy Gamow Shell Model framework to dipolar anions,
are gratefully acknowledged. This work was
supported by
the U.S. Department of Energy under
Contract No.\ DE-FG02-96ER40963.
\end{acknowledgments}

\bibliographystyle{apsrev}
\bibliography{ANIONS}

\end{document}